\documentclass[prb,preprint,showpacs,preprintnumbers,amsmath,amssymb]{revtex4}
\usepackage{bm}
\usepackage{graphicx}

\begin{document}
\title{FIELD-INDUCED SPIN EXCITATIONS IN RASHBA-DRESSELHAUS TWO-DIMENSIONAL
ELECTRON SYSTEMS PROBED BY SURFACE ACOUSTIC WAVES}

\author{P. Kleinert}
\affiliation{Paul-Drude-Intitut f\"ur Festk\"orperelektronik,
Hausvogteiplatz 5-7, 10117 Berlin, Germany}
\author{V.V. Bryksin}
\affiliation{A.F. Ioffe Physical Technical Institute,
Politekhnicheskaya 26, 194021 St. Petersburg, Russia}
\date{\today}

\begin{abstract}
A spin-rotation symmetry in spin-orbit coupled two-dimensional
electron systems gives rise to a long-lived spin excitation that
is robust against short-range impurity scattering. The influence
of a constant in-plane electric field on this persistent spin
helix is studied. To probe the field-induced eigen-modes of the
spin-charge coupled system, a surface acoustic wave is exploited
that provides the wave-vector for resonant excitation. The
approach takes advantage of methods worked out in the field of
space-charge waves. Sharp resonances in the field dependence of
the in-plane and out-of-plane magnetization are identified.
\end{abstract}


\maketitle

\section{Introduction}
The perspectives of exploiting electron spin for information
processing in spintronic devices stimulated basic research of
spin-related phenomena.\cite{RMP_323} In order to take advantage
of traditional semiconductor technologies, the exclusive
application of electric fields to generate and manipulate a
nonequilibrium spin density is of particular interest. In this
respect, the spin-orbit interaction (SOI) provides an attractive
mechanism for the pure electric manipulation of moving spins in
nonmagnetic semiconductors. Unfortunately, the very same SOI
depends on the carrier momentum, which is randomized by elastic
and inelastic scattering. This momentum dependence of the coupling
leads to the dominant spin relaxation and dephasing in III-V
semiconductors known as Dyakonov-Perel relaxation
mechanism.\cite{Dyakonov} In general, however, details of spin
dephasing depend on the involved scattering mechanisms, the band
structure, and the crystal orientation (see Ref. \cite{JAP_073702}
and references therein).

An sizable anisotropy of the spin relaxation has been
theoretically predicted~\cite{PRB_15582,JPC_R271} for
semiconductor heterostructures grown along the [001] axis.
According to these studies, a large spin-relaxation anisotropy
should exist for the in-plane spin polarization. Moreover, under
the condition that the strengths of Rashba ($\alpha$) and
Dresselhaus ($\beta$) spin-orbit coupling constants become equal
($\alpha=\beta$), the anisotropy should be maximal. Based on this
observation a proposal for a novel spin-transistor device
\cite{PRL_146801} was put forward. Recent experimental data
provided strong evidence for a huge in-plane anisotropy of the
spin dephasing time in [001] GaAs/Al$_{x}$Ga$_{1-x}$As quantum
wells.\cite{PRB_033305,APL_112111,PRB_073309} The efficient
suppression of relaxation for spins aligned along the [110]
orientation in heterostructures with equal Rashba and Dresselhaus
SOI strengths is an important result that certainly stimulates
further progress in spintronics.

Other promising structures for spintronic applications are quantum
well samples grown along the [110] direction. Very long
spin-relaxation times on the order of nanoseconds were
experimentally demonstrated for the out-of plane spin polarization
in such heterostructures.\cite{PRL_4196,PRL_036603} This effect is
caused by the absence of Dyakonov-Perel spin relaxation in the
particular crystal direction.

The common physical origin for weak spin relaxation in [110]
quantum wells as well as in [001] heterostructures with
$\alpha=\beta$ is the presence of an spin-rotation symmetry in the
two-dimensional electron gas (2DEG).\cite{PRL_236601} This
symmetry is robust and leads to an infinite spin lifetime and a
persistent spin helix for an idealized model. The theoretical
explanation, which has recently been confirmed by
experiment,\cite{PRL_076604W} will initiate a number of further
studies in this field. It is the aim of this paper to contribute
to this development by treating weakly damped eigen-excitations of
the coupled spin-charge system subject to a constant electric
field. Our approach bears some similarities with a traditional
field in solid state physics namely the study of space-charge
waves in crystals.\cite{BuchPetrov1} The electric-field-mediated
excitations of the coupled spin-charge system are probed by a
simulated experimental set up, which provides the wave vector to
resonantly excite internal eigen-modes of the biased SOI coupled
2DEG. We focus on excitation mechanisms that exploit surface
acoustic waves.

\section{Basic Theory}
Our model is based on a Hamiltonian that describes conduction band
electrons in a III-V semiconductor quantum well grown in the [001]
direction, which is used to be the $z$ axis of the coordinate
system
\begin{equation}
H_0=\frac{\hbar^2\bm{k}^2}{2m}+\hbar\bm{\omega}_{\bm{k}}\cdot\bm{\sigma}.
\label{Hamil}
\end{equation}
The effective electron mass, the in-plane wave vector, and the
Pauli matrices are denoted by $m$, $\bm{k}$, and $\sigma_{i}$
($i=x,y,z$), respectively. The precession frequencies
$\bm{\omega}_{\bm{k}}$ are due to the SOI, which contain both
linear Rashba and Dresselhaus contributions
\begin{equation}
\bm{\omega}_{\bm{k}}^{R}=\frac{\alpha}{\hbar}(k_y,-k_x),\quad
\bm{\omega}_{\bm{k}}^{D}=\frac{\beta}{\hbar}(k_x,-k_y),
\end{equation}
with $\alpha$, $\beta$ being the respective spin-orbit coupling
constants. These vectors are also interpreted as a
momentum-dependent internal magnetic field that acts on the
carrier motion. The neglected cubic terms to the frequency vector
$\bm{\omega}_{\bm{k}}$ play only a secondary role in our study.
Throughout this work, it is assumed that the spin-orbit coupling
energy is much smaller than the Fermi energy $\varepsilon_F$. The
model is complemented by the inclusion of elastic scattering on a
short-range impurity potential characterized by the elastic
scattering time $\tau$.

To elucidate the peculiarities of the spin-orbit coupled 2DEG, the
coordinate system is rotated so that the [110] direction becomes a
coordinate axis ($k_{\pm}=(k_x\pm k_y)/\sqrt{2}\rightarrow
r_{\pm}=(r_x\pm r_y)/\sqrt{2}$). Along the spatial $r_{+}$ axis, a
constant electric field $E_0$ is applied. In addition, a surface
acoustic wave (SAW) with the wave number $K_{SAW}$ and frequency
$\Omega=v_{SAW} K_{SAW}$ ($v_{SAW}$ is the sound velocity) should
propagate in the same direction giving rise to an internal
electric field. The total electric field $E_{+}$ is expressed by
\begin{equation}
E_{+}(r_{+},t)=E_0+E_{SAW}\cos(K_{SAW}r_{+}-\Omega t)+\delta
E(r_{+},t),\label{Feld}
\end{equation}
is self-consistently calculated by taking into account the Poisson
equation
\begin{equation}
\frac{\partial\delta E(r_{+},t)}{\partial r_{+}}=\frac{4\pi
e}{\varepsilon}(F-n),\label{Pois}
\end{equation}
with $F$ and $n$ denoting the nonequilibrium and initially
homogeneous carrier density, respectively (the components of the
spin-density matrix $F_{\mu}$, ($\mu =1,2,3,4$) are defined as in
Ref.~\cite{PRB_205326}). $\varepsilon$ is the dielectric constant.
Due to the excitation conditions in Eqs.~(\ref{Feld}) and
(\ref{Pois}), all quantities are independent of the spatial
coordinate $r_{-}$ for an infinite 2DEG. Along the $r_{+}$ axis,
periodic boundary conditions are assumed.

All electric field components act both on the charge and spin
degrees of freedom leading to specific field-induced eigen-modes
of the spin-charge-coupled system. The appearance and excitation
of these modes is treated by drift-diffusion equations that are
obtained from a kinetic theory for the spin-density matrix. For
the given orientation of the constant electric field $E_0$ and the
SAW field, the set of equations decouples and takes the form
\cite{PRB_205326}
\begin{equation}
\left[\frac{\partial}{\partial t}-D\frac{\partial^2}{\partial
r_{+}^2} \right]F+\mu\frac{\partial E_{+}F}{\partial r_{+}}
-\frac{\hbar}{2m}K_{+}\frac{\partial F_{-}}{\partial r_{+}}
=0,\label{eq_F}
\end{equation}
\begin{equation}
\left[\frac{\partial}{\partial t}-D\frac{\partial^2}{\partial
r_{+}^2}+\frac{2}{\tau_{s-}} \right]F_{-}+\mu\frac{\partial
E_{+}F_{-}}{\partial r_{+}}
-\frac{2}{\tau_{s-}}\frac{F_{-}^{(0)}}{nE_0}\left[
E_{+}-\frac{2D}{\mu}\frac{\partial}{\partial r_{+}}\right]F=0,
\label{eq_Fm}
\end{equation}
\begin{equation}
\left[\frac{\partial}{\partial t}-D\frac{\partial^2}{\partial
r_{+}^2}+\frac{2}{\tau_{s+}} \right]F_{+}+\mu\frac{\partial
E_{+}F_{+}}{\partial r_{+}}-K_{+}\left(\mu
E_{+}-2D\frac{\partial}{\partial r_{+}}\right)F_{z}=0,
\label{eq_Fp}
\end{equation}
\begin{equation}
\left[\frac{\partial}{\partial t}-D\frac{\partial^2}{\partial
r_{+}^2}+\frac{2}{\tau_{s}} \right]F_{z}+\mu\frac{\partial
E_{+}F_{z}}{\partial r_{+}}+K_{+}\left(\mu E_{+}-2D\frac{\partial
}{\partial r_{+}}\right)F_{+}=G_z, \label{eq_Fz}
\end{equation}
where $F_{\pm}$ and $F_z$ denote the in-plane and out-of plane
nonequilibrium spin densities, respectively. $D$ is the diffusion
coefficient, $\mu$ the carrier mobility, and
$K_{+}=2m(\alpha+\beta)/\hbar^2$ a wave number constructed from
the spin-orbit coupling constants. In addition, we introduced the
spin-relaxation times
\begin{equation}
\frac{1}{\tau_s}=4DK^2,\quad
\frac{1}{\tau_{s+}}=\frac{\cos^2\varphi}{\tau_{s}},\quad
\frac{1}{\tau_{s-}}=\frac{\sin^2\varphi}{\tau_{s}}, \label{spints}
\end{equation}
with $K=m\sqrt{\alpha^2+\beta^2}/\hbar^2$. The angle $\varphi$ is
used to quantify the coupling constants according to
$\alpha/\sqrt{\alpha^2+\beta^2}=\cos(\varphi+\pi/4)$ and
$\beta/\sqrt{\alpha^2+\beta^2}=\sin(\varphi+\pi/4)$. For the
limiting cases of a pure Rashba and Dresselhaus model, we have
$\varphi=-\pi/4$ and $\pi/4$, respectively. When the SAW field is
switched off, the applied constant electric field $E_0$ induces a
homogeneous in-plane magnetization $F_{-}^{(0)}$ that is
calculated from
\begin{equation}
F_{-}^{(0)}=\sqrt{2}\frac{\cos^2\varphi}{\sin\varphi}\hbar K\mu
E_0\frac{dn}{d\varepsilon_F}.
\end{equation}
Taking into account the Einstein relation $\mu=eD d\ln
n/d\varepsilon_F$, this magnetization is expressed by the result
$F_{-}^{(0)}=\hbar K \mu E_0 dn/d\varepsilon_F$ (with
$K=m\alpha/\hbar^2$) for a Rashba model ($\beta=0$) that was
derived by Edelstein.~\cite{Edelstein} When the SOI coupling
strengths become equal ($\alpha=\beta$), the spin-relaxation time
$\tau_{s-}$ diverges so that the result for $F_{-}^{(0)}$ strongly
depends on the initial condition and/or additional spin relaxation
mechanisms quantified by the scattering time $\tau_{sp}$
($\tau_{s-}^{-1}\rightarrow \tau_{s-}^{-1}+\tau_{sp}^{-1}$). The
divergency is also circumvented in the ac
response.\cite{PRB_205326,IJMPB_4937} In all cases, the
field-induced in-plane magnetization disappears for
$\alpha=\beta$. In Eq.~(\ref{eq_Fz}), $G_z$ describes the uniform
generation of an out-of plane spin polarization by optical means
or by the application of a perpendicular constant magnetic field.

From Eqs.~(\ref{Pois}) and (\ref{eq_F}), an expression for the
total time-dependent charge current density $I(t)$ is obtained
\begin{equation}
j(r_{+},t)+\frac{\varepsilon}{4\pi}\frac{\partial \delta
E}{\partial t}=I(t), \quad j(r_{+},t)=e\mu E_{+}F-eD\frac{\partial
F}{\partial r_{+}}-\sqrt{2}\frac{\hbar K}{m}\cos\varphi eF_{-},
\label{jdef}
\end{equation}
which completes the set of basic equations necessary to study
electric-field-mediated effects in the spin-charge coupled 2DEG.
The drift-diffusion Eqs.~(\ref{eq_F}) to (\ref{eq_Fz}) decouple
into two sets of equations for the components $F$, $F_{-}$ and
$F_{z}$, $F_{+}$. Accordingly, the spin-charge coupling prescribed
by Eqs.~(\ref{eq_F}) and (\ref{eq_Fm}) is treated in the next
Section, whereas Eqs.~(\ref{eq_Fp}) and (\ref{eq_Fz}) are solved
in Section 4.

\section{Spin-charge coupling}
The solution of Eqs.~(\ref{eq_F}) to (\ref{eq_Fz}) is facilitated
by the fact that all quantities depend only on
$z=K_{SAW}r_{+}-\Omega t$. The derivative with respect to $z$ is
denoted by a prime. Introducing the dimensionless functions
$f=F/n$, $f_{\pm}=F_{\pm}/n$, and $f_z=F_z/n$ and the parameters
\begin{equation}
Y_{+}=\frac{E_{+}}{E_0}, \quad Y=\frac{\delta E}{E_0}, \quad
\Omega_E=\mu E_0 K_{SAW},\quad \frac{1}{\tau_M}=\frac{4\pi
e}{\varepsilon}\mu n, \quad \Lambda=\frac{DK_{SAW}}{\mu E_0},
\end{equation}
we obtain the following set of coupled ordinary differential
equations
\begin{equation}
-\Omega f_{-}^{\prime}-\Lambda\Omega_E
f_{-}^{\prime\prime}+\frac{2}{\tau_{s-}}f_{-}
+\Omega_E\left(Y_{+}f_{-}
\right)^{\prime}+\frac{2}{\tau_{s-}}f_{-}^{(0)}\left(Y_{+}f-2\Lambda
f^{\prime} \right)=0, \label{dichte1}
\end{equation}
\begin{equation}
-\Omega\tau_{M}Y^{\prime}+Y_{+}f-\Lambda
f^{\prime}-\sqrt{2}\frac{\hbar K}{m\mu E_0}\cos\varphi
f_{-}=\frac{I}{j_0},\quad \Omega_E\tau_{M}Y^{\prime}=f-1.
\label{dichte2}
\end{equation}
These equations are solved by applying perturbation theory with
respect to the SAW electric field $E_{SAW}$. When the SAW field is
absent ($E_{SAW}=0$), all quantities are independent of $z$ and we
obtain for the constant charge current density
\begin{equation}
I_0=j_0\left(1-\sqrt{2}\frac{\hbar K}{m\mu E_0}\cos\varphi
f_{-}^{(0)} \right), \label{I0}
\end{equation}
with $j_0=en\mu E_0$. The spin-induced contribution on the
right-hand side of this equation results from the homogeneous
in-plane magnetization $f_{-}^{(0)}$ that vanishes for a system
with equal Rashba and Dresselhaus coupling constants
($\alpha=\beta$).

To proceed, we calculate the stationary current contribution
$\delta I$ associated with the SAW electric field. This
time-independent current contribution arises from carriers that
are driven by the SAW. A similar current can also be generated by
a moving optical lattice in a photorefractive crystal. As a
periodic boundary condition along the $r_{+}$ axis is assumed, the
solution is searched for in the form of a Fourier series
$f(z)=n+\delta f(z)$, $\delta f(z)=\sum_{p}\exp(ipz)\delta f(p)$.
Taking into account the Fourier transformed version of
Eq.(\ref{jdef})
\begin{equation}
\frac{I}{j_0}\delta_{p,0}=-ip\Omega\tau_MY(p)+\sum
\limits_{p^{\prime}}Y_{+}(p-p^{\prime})f(p^{\prime}) -ip\Lambda
f(p)-\sqrt{2}\frac{\hbar K}{m\mu E_0}\cos\varphi f_{-}(p),
\end{equation}
we obtain
\begin{equation}
\frac{\delta I}{j_0}=\frac{Y_{SAW}}{2}\left[1-\sqrt{2}\frac{\hbar
K}{m\mu E_0}\cos\varphi f_{-}^{(0)} \right]\left[\delta
f(1)+\delta f(-1)\right], \label{ee}
\end{equation}
with $Y_{SAW}=E_{SAW}/E_0$. The nonequilibrium fluctuation of the
charge density $\delta f(1)$ for $p=1$ is expressed by the
variation of the internal electric field $\delta Y(1)$ via $\delta
Y(1)=-i\delta f(1)/(\Omega_E\tau_M)$. The latter quantity is
calculated from the Fourier transformed Eqs.~(\ref{dichte1}) and
(\ref{dichte2}), which are given by
\begin{eqnarray}
&&
\left[-ip(\Omega-\Omega_E+ip\Lambda\Omega_E)+\frac{2}{\tau_{s-}}
\right]\delta f_{-}(p)\label{e1}\\
&&+f_{-}^{(0)}\left[ip\Omega_E+\frac{2}{\tau_{s-}}\left(1+ip\Omega_E\tau_M
\right) \right]\delta
Y(p)+4p^2\frac{\Lambda}{\tau_{s-}}f_{-}^{(0)}\delta
Y(p)\nonumber\\
&&=-\frac{1}{\tau_{s-}}f_{-}^{(0)}Y_{SAW}(\delta_{p,1}+\delta_{p,-1})-\frac{i\Omega_E}{2}
f_{-}^{(0)}Y_{SAW}(\delta_{p,1}-\delta_{p,-1}),\nonumber
\end{eqnarray}
\begin{eqnarray}
&&\left[1-ip\tau_M(\Omega-\Omega_E(1-ip\Lambda)) \right]\delta
Y(p)-\sqrt{2}\frac{\hbar K}{m\mu E_0}\cos\varphi\delta
f_{-}(p)\nonumber\\
&&=-\frac{1}{2}Y_{SAW}(\delta_{p,1}+\delta_{p,-1}). \label{e2}
\end{eqnarray}
Even at the absence of SOI, the SAW field alone gives rise to a
charge current that is easily calculated from Eqs.~(\ref{e1}),
(\ref{e2}), and (\ref{ee})
\begin{equation}
\left(\frac{\delta I}{j_0}\right)_1=\frac{Y_{SAW}^2\Omega_E}{2}
\frac{\Omega-\Omega_E}{(\Omega-\Omega_E)^2+(\tau_M^{-1}+DK_{SAW}^2)^2}.
\label{jSAW}
\end{equation}
This current contribution is due to space-charge waves, the
dispersion relation of which is obtained from the poles in
Eq.~(\ref{jSAW})
\begin{equation}
\Omega=\mu E_0K_{SAW}- i\left(\frac{1}{\tau_M}+DK_{SAW}^2 \right),
\end{equation}
where $\tau_M$ denotes the Maxwellian relaxation time. The
physical origin of this mode, which has been thoroughly studied in
the field of photorefractive materials~\cite{BuchPetrov1}, are
oscillations of the free electron gas. In the same way as the
current density $j_0$ has a spin complement in Eq.~(\ref{I0}), the
well-known SAW-induced charge current in Eq.~(\ref{jSAW}) has a
related spin contribution, which is also calculated from
Eqs.~(\ref{ee}) and (\ref{e1}), (\ref{e2}) with the result
\begin{eqnarray}
&&\left(\frac{\delta
I}{j_0}\right)_2=\frac{\hbar}{4m}\frac{K_{+}K_{SAW}f_{-}^{(0)}Y_{SAW}^2}
{\left[(\Omega-\Omega_E)^2+\left(\tau_M^{-1}+DK_{SAW}^2 \right)^2
\right]\left[\left(\Omega-\Omega_E
\right)^2+\left(2/\tau_{s-}+DK_{SAW}^2 \right)^2
\right]}\nonumber\\
&&\times\Biggl\{(\Omega-\Omega_E)
\left(\frac{2}{\tau_{s-}}+DK_{SAW}^2 \right)
\left(\frac{2}{\tau_{s-}}-\frac{1}{\tau_M} \right)\label{jend}\\
&& +\Omega\left[\left(\frac{1}{\tau_{M}}+DK_{SAW}^2
\right)\left(\frac{2}{\tau_{s-}}+DK_{SAW}^2 \right)-
\left(\Omega-\Omega_E \right)^2 \right]\Biggl\}.\nonumber
\end{eqnarray}
The SAW-induced electronic current in Eq.~(\ref{jSAW}) is
independent of the spin-degree of freedom and disappears at the
resonance frequency $\Omega=\Omega_E$ of space-charge waves. This
result has to be compared with Eq.~(\ref{jend}) that holds for the
spin complement of the SAW-induced charge current, which may
exhibit a sharp resonance at $\Omega=\Omega_E$. Besides
space-charge excitations that also appear in Eq.~(\ref{jSAW}),
there exists a new eigen-mode, in which $\tau_M^{-1}$ is replaced
by the spin-scattering rate $2/\tau_{s-}$ that vanishes for
$\alpha=\beta$. This spin-mediated mode has the character of a
space-charge wave with an infinite Maxwellian relaxation time (for
$\alpha=\beta$). In experiment, the SAW related stationary charge
currents in Eqs.~(\ref{jSAW}) and (\ref{jend}) can be
distinguished by their qualitative different dependence on the
electric field $E_0$.

\section{Spin-spin coupling}
For the considered set-up not all four components of the
spin-density matrix couple to each other. According to the special
field orientation, the coupling between the spin components $F_z$
and $F_{+}$ is separated from the charge density $F$ and the
in-plane spin component $F_{-}$. The related drift-diffusion
Eqs.~(\ref{eq_Fp}) and (\ref{eq_Fz}) take the dimensionless form
\begin{equation}
-\Omega f_{+}^{\prime}-\Lambda\Omega_E
f_{+}^{\prime\prime}+\frac{2}{\tau_{s+}}f_{+}
+\Omega_E(Y_{+}f_{+})^{\prime}-\frac{K_{+}\Omega_E}{K_{SAW}}Y_{+}f_z
+2\frac{K_{+}\Omega_E\Lambda}{K_{SAW}}f_z^{\prime}=0, \label{b1}
\end{equation}
\begin{equation}
-\Omega f_{z}^{\prime}-\Lambda\Omega_E
f_{z}^{\prime\prime}+\frac{2}{\tau_{s}}f_{z}
+\Omega_E(Y_{+}f_{z})^{\prime}+\frac{K_{+}\Omega_E}{K_{SAW}}Y_{+}f_{+}
-2\frac{K_{+}\Omega_E\Lambda}{K_{SAW}}f_{+}^{\prime}=\frac{G_z}{n},
\label{b2}
\end{equation}
where $G_z$ denotes the source of an out-of-plane spin
polarization. Without the application of the SAW field, we obtain
the steady-state solution
\begin{equation}
f_{z}^{(0)}=\frac{G_{z}/n}{2/\tau_{s}+(\mu E_0)^2/D},\quad
f_{+}^{(0)}=-\frac{\mu E_0}{DK_{+}}f_{z}^{(0)}.\label{fz0}
\end{equation}
The out-of-plane spin polarization $f_z^{(0)}$ is suppressed by
the electric field.\cite{Kalevich} By determining this suppression
in experiment, the spin-relaxation time $\tau_s$ can be determined
in a way similar to the well established experimental set up that
is based on the Hanle effect (c.f., for instance, Ref.
\cite{PRB_033305}). Furthermore, the electric field $E_0$ induces
an in-plane spin polarization $f_{+}^{(0)}$.

As in Section 3, the effect of the SAW is treated by a
perturbation expansion $f_z=f_z^{(0)}+\delta f_z$ and
$f_{+}=f_{+}^{(0)}+\delta f_{+}$. The Fourier transformed basic
Eqs.~(\ref{b1}) and (\ref{b2}) are easily solved, when non-linear
perturbations are neglected. The solution for the lowest-order
Fourier components is given in the Appendix.

To determine the steady-state spin polarization under the mutual
influence of a constant electric field and the SAW field, we focus
on the most interesting case $\alpha=\beta$ and assume
$\Omega_E\tau_M\gg 1$. Under these conditions, the solution given
by Eqs.~(\ref{A3}) to (\ref{A5}) simplifies and takes the form
\begin{equation}
\frac{\delta
f_{+}(0)}{f_{+}^{(0)}}=-\frac{Y_{SAW}^2\Omega_E}{2(1+\Lambda_{+}^2)}{\rm
Re}\Biggl\{\frac{\Omega_E(1-q^2)Z_1
+\Omega(1-\Lambda_{+}^2-2iq\Lambda_{+})}{(\Omega-\Omega_{+})(\Omega-\Omega_{-})}
\Biggl\},
\end{equation}
\begin{equation}
\frac{\delta
f_{z}(0)}{f_{z}^{(0)}}=\frac{Y_{SAW}^2\Omega_E}{2\Lambda(1+\Lambda_{+}^2)}{\rm
Re}\Biggl\{\frac{\Omega_E(1-q^2)Z_2
+i\Omega(1-\Lambda_{+}^2+2iq\Lambda)}{(\Omega-\Omega_{+})(\Omega-\Omega_{-})}
\Biggl\},
\end{equation}
with the shorthand notations
$Z_1=3\Lambda_{+}-1+i\Lambda(1-\Lambda_{+}^2)$,
$Z_2=\Lambda(3-2i\Lambda +\Lambda_{+}^2)$, $q=K_{+}/K_{SAW}$, and
$\Lambda_{+}=q\Lambda$. The eigen-frequencies of the spin
excitations are expressed by
\begin{equation}
\Omega_{\pm}=\mu E_0(K_{SAW}\pm K_{+})-iD(K_{SAW}\pm K_{+})^2.
\label{disper}
\end{equation}
The appearance of resonances at $\Omega=\Omega_{\pm}$ that occur
in the in-plane and out-of-plane field-induced spin polarization
is the most remarkable result of this paper. Due to the
spin-rotation symmetry, a soft mode develops in the system that
can be probed by a SAW, which provides the wave vector $K_{SAW}$
for the resonant excitation. The low-frequency mode
$\Omega\rightarrow 0$ becomes increasingly undamped for
$K_{SAW}\rightarrow \pm K_{\pm}$. As a remnant of these coherent
long-lived spin excitations, a weakly damped resonance appears,
when the applied constant electric field $E_0$ satisfies the
condition $\Omega=v_{SAW}K_{SAW}=\mu E_0(K_{SAW}-K_{+})$.

\begin{figure}
\begin{minipage}[l]{0.47\textwidth}
\includegraphics[width=2.5in]{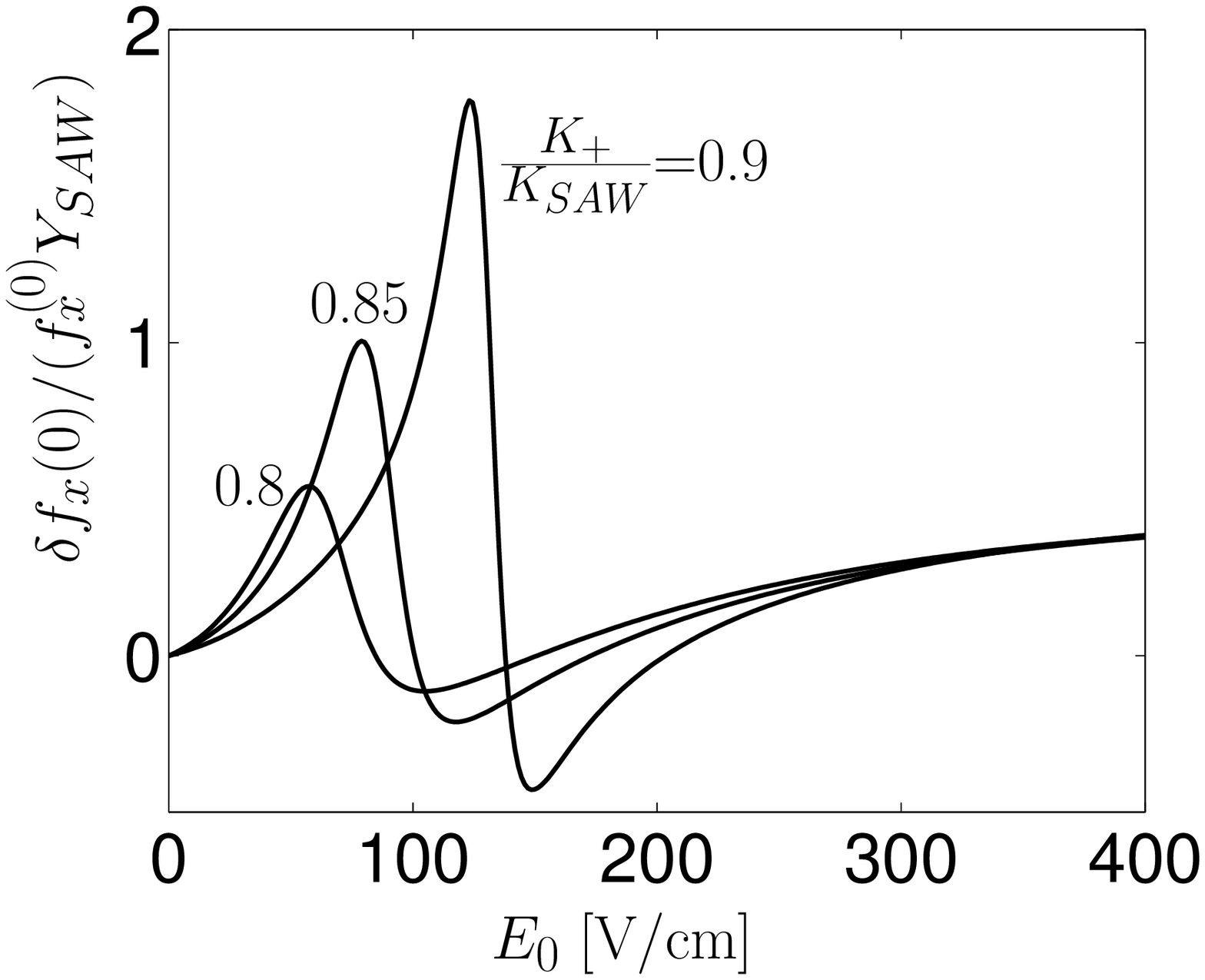}%
\caption{\small SAW-induced stationary in-plane spin polarization
$\delta f_x(0)$ referred to $f_x^{(0)}Y_{SAW}$ for
$\lambda_{SAW}=5.6$ $\mu$m, $\mu=2.3\,10^{4}$ cm$^{2}$/Vs, and
$D=24$ cm$^{2}$/s. Results are shown for $K_{+}/K_{SAW}=$ 0.8,
0.85, and 0.9.} \label{fig1}
\end{minipage}%
\hspace{0.5cm}
\begin{minipage}[r]{0.47\textwidth}
\includegraphics[width=2.5in]{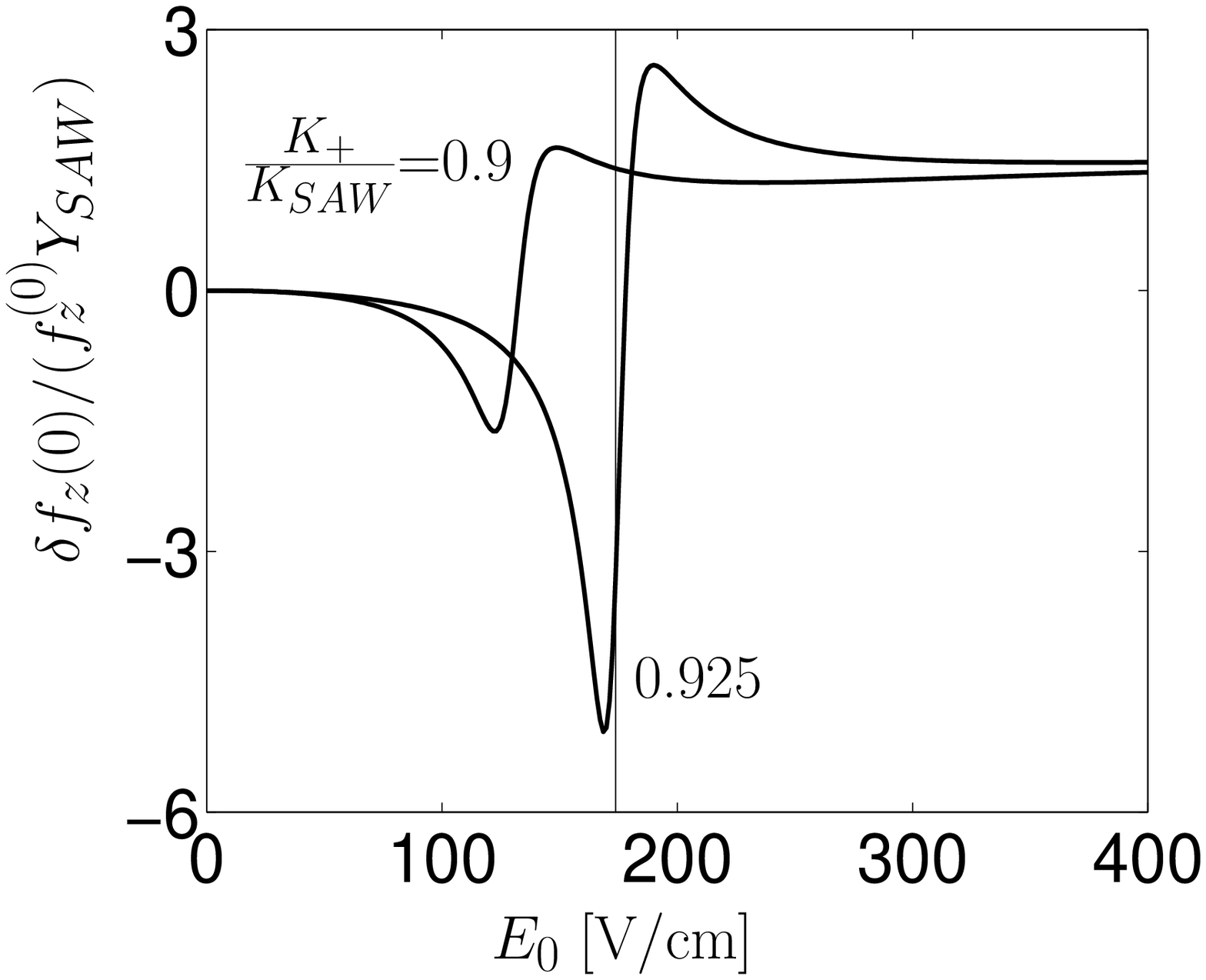}%
\caption{\small SAW-induced stationary out-of-plane spin
polarization $\delta f_z(0)$ referred to $f_z^{(0)}Y_{SAW}$ for
$\lambda_{SAW}=5.6$ $\mu$m, $\mu=2.3\,10^{4}$ cm$^{2}$/Vs, and
$D=24$ cm$^{2}$/s. Results are shown for $K_{+}/K_{SAW}=$ 0.9 and
0.925.} \label{fig2}
\end{minipage}
\end{figure}

Numerical examples are shown in Figs. \ref{fig1} and \ref{fig2}
for the SAW-induced
in-plane and out-of-plane spin polarization. The parameters used
in the calculation refer to GaAs quantum
wells.\cite{PRL_036603,PRL_047602} It is to be noted that in the
displayed relative spin polarizations, the electric field $E_0$
enters via the field-dependent quantities $f_x^{(0)}$ and
$f_z^{(0)}$ calculated from Eq.~(\ref{fz0}). Both components of
the magnetization exhibit a sharp resonance in the electric field
dependence at $\mu E_0=v_{SAW}/(1-K_{+}/K_{SAW})$ as indicated by
the vertical line in Fig.~\ref{fig2}. With increasing ratio
$K_{+}/K_{SAW}$, the resonance is shifted to higher field
strengths and becomes more pronounced. Under the condition
$E_{SAW}\approx E_0$ and at the resonance, the out-of-plane spin
polarization may significantly exceed the spin generation
$f_z^{(0)}$ given in Eq.~(\ref{fz0}). As the suppression of
$f_z^{(0)}$ at high electric fields can be compensated by an
appropriate in-plane magnetic field~\cite{PRB_075340}, the
experimental demonstration of the effect should be feasible. Its
most salient feature is the strong variation of the field-induced
magnetization in the vicinity of the resonance field strength.
This switching of the magnetization caused by an applied constant
electric field could be useful for future spintronic device
applications.

\section{Summary}
Recently, long-range transport of spins has been experimentally
demonstrated in [001] GaAs quantum wells with balanced Rashba and
Dresselhaus terms as well as in Dresselhaus [110] quantum wells.
The idealized model that describes both systems exhibits an exact
spin rotation symmetry with an associated soft mode and a
persistent spin helix. This long-lived spin coherence state is
transferred to a field-dependent eigenmode of the system, when a
constant in-plane electric field is applied. To resonantly probe
the field-induced excitation, the required wave-vector must be
provided by an appropriate experimental set up. Similar to the
well established physics of space-charge waves one can produce
optical gratings for that purpose. In this paper, we treated an
excitation mechanism via an acoustic wave propagating along a
[001] GaAs quantum well, in which both Rashba and Dresselhaus SOI
exist. Due to the spin-charge coupling, specific eigenmodes are
identified both in the charge current density and the in-plane and
out-of-plane spin polarizations. The SAW-induced change of the
stationary charge current is composed of two contributions. The
first one is independent of the SOI and exhibits a pole that is
due to oscillations of the free carrier density. In the second
contribution, which is related to the field-induced homogeneous
spin accumulation, a new eigenmode appears with the dispersion
relation $\Omega=\mu E_0 K_{SAW}- (DK_{SAW}^2+2/\tau_{s-})$. In
this equation, the spin-scattering time $\tau_{s-}$ (which
diverges in the case $\alpha=\beta$) takes over the role of the
Maxwellian relaxation time $\tau_M$ that is responsible for the
damping of pure charge oscillations.

Characteristic field-dependent eigenmodes appear also in the
SAW-induced components of the magnetization. Most interesting is
the case, when the Rashba and Dresselhaus coupling strengths
become equal. According to the dispersion relation
$\Omega=\Omega_{-}$, a undamped soft mode occurs, when the
wave-vector $K_{SAW}$ of the SAW approaches the shifting vector
$K_{+}=2m(\alpha+\beta)/\hbar^2$. As a remnant of this persistent
spin helix, a sharp resonance occurs in the electric-field
dependence of the in-plane and out-of-plane magnetization at $\mu
E_0=v_{SAW}/(1-K_{+}/K_{SAW})$. In the vicinity of this resonance,
the spin polarization drastically changes with a slight variation
of the electric field strength. It is a challenge to
experimentally verify this field-induced switching of
magnetization in a 2DEG with balanced Rashba and Dresselhaus SOI.



\appendix{Solution of kinetic equations}
In this Section, the coupled Eqs.~(\ref{b1}) and (\ref{b2}) are
solved by a perturbation approach with respect to the SAW electric
field $Y_{SAW}$. First, the equations are expressed in terms of a
Fourier series, which leads to the result
\begin{eqnarray}
&&\left[-ip\Omega+p^2\Lambda\Omega_E+\frac{2}{\tau_{s+}}+ip\Omega_E
\right]\delta f_{+}(p)+\frac{ipY_{SAW}\Omega_E}{2}\left[\delta
f_{+}(p+1)+\delta f_{+}(p-1)\right]\nonumber\\
&&-\frac{K_{+}\Omega_E}{K_{SAW}}(1-2ip\Lambda)\delta
f_z(p)-\frac{K_{+}Y_{SAW}\Omega_E}{2K_{SAW}}\left[\delta
f_{z}(p+1)+\delta
f_{z}(p-1)\right]\label{A1}\\
&&=f_{+}^{0}\Omega_E\left[\frac{Y_{SAW}}{2i}(\delta_{p,1}-\delta_{p,-1})
-ip Y(p)
\right]+\frac{K_{+}f_z^{0}\Omega_E}{K_{SAW}}\left[\frac{Y_{SAW}}{2}(\delta_{p,1}+\delta_{p,-1})
+Y(p) \right],\nonumber
\end{eqnarray}
\begin{eqnarray}
&&\left[-ip\Omega+p^2\Lambda\Omega_E+\frac{2}{\tau_{s}}+ip\Omega_E
\right]\delta f_{z}(p)+\frac{ipY_{SAW}\Omega_E}{2}\left[\delta
f_{z}(p+1)+\delta f_{z}(p-1)\right]\nonumber\\
&&+\frac{K_{+}\Omega_E}{K_{SAW}}(1-2ip\Lambda)\delta
f_{+}(p)+\frac{K_{+}Y_{SAW}\Omega_E}{2K_{SAW}}\left[\delta
f_{+}(p+1)+\delta
f_{+}(p-1)\right]\label{A2}\\
&&=f_{+}^{0}\Omega_E\left[\frac{Y_{SAW}}{2i}(\delta_{p,1}-\delta_{p,-1})
-ip Y(p)
\right]-\frac{K_{+}f_{+}^{0}\Omega_E}{K_{SAW}}\left[\frac{Y_{SAW}}{2}(\delta_{p,1}+\delta_{p,-1})
-Y(p) \right].\nonumber
\end{eqnarray}
To calculate the steady-state solution, the equations for the
$p=0$ components $\delta f_{+}(0)$ and $\delta f_z(0)$ are solved.
We obtain
\begin{equation}
\delta
f_{+}(0)=\frac{Y_{SAW}}{1+2(\alpha^2+\beta^2)\Lambda_{+}^2/(\alpha+\beta)^2}{\rm
Re}\left[\Lambda_{+}\delta f_{z}(1)-\delta f_{+}(1)
\right],\label{A3}
\end{equation}
\begin{equation}
\delta
f_{z}(0)=-\frac{Y_{SAW}}{1+2(\alpha^2+\beta^2)\Lambda_{+}^2/(\alpha+\beta)^2}{\rm
Re}\left[\Lambda_{+}\delta f_{+}(1)+\delta f_{z}(1)
\right].\label{A4}
\end{equation}
In a second step, the Fourier components $\delta f_{+}(1)$ and
$\delta f_z(1)$ have to be calculated. Disregarding nonlinear
contributions, the remaining equations are easily solved with the
result
\begin{equation}
\delta f_{+}(1)=\frac{c_1A_{s}-c_2B}{A_{s}A_{s+}+B^2},\quad \delta
f_{z}(1)=-\frac{c_2A_{s+}+c_1B}{A_{s}A_{s+}+B^2},\label{A5}
\end{equation}
$$
c_1=\left(Y(1)+\frac{Y_{SAW}}{2}\right)\left(-if_{+}^{0}+\frac{K_{+}f_{+}^{0}}{K_{SAW}}
\right),\quad
c_2=\left(Y(1)+\frac{Y_{SAW}}{2}\right)\left(if_{z}^{0}+\frac{K_{+}f_{+}^{0}}{K_{SAW}}
\right),
$$
$$
A_{s}=\frac{2}{\Omega_{E}\tau_{s}}+i(1-i\Lambda-\frac{\Omega}{\Omega_E}),\quad
A_{s+}=\frac{2}{\Omega_{E}\tau_{s+}}+i(1-i\Lambda-\frac{\Omega}{\Omega_E}),\quad
B=\frac{K_{+}}{K_{SAW}}(1-2i\Lambda).
$$
The SAW-induced stationary magnetization of the biased 2DEG is
expressed by Eqs.~(\ref{A3}) to (\ref{A5}).


\end{document}